\documentclass[preprint,aps,prl,superscriptaddress,amsmath,amssymb,bibnotes,showkeys]{revtex4-1}
\usepackage{graphicx}
\usepackage{amsmath}
\usepackage{dcolumn}
\usepackage{bm}
\usepackage{braket} 
\usepackage{bbold}
\usepackage{textcomp}
\usepackage{multirow}
\usepackage{url}
\usepackage{endnotes}
\usepackage{tabularx}
\usepackage{color}
\usepackage{soul}
\usepackage[
 colorlinks=true,
 urlcolor=blue,
 citecolor=blue,
 linkcolor=blue,
 bookmarks=false,
 pdfstartview={FitH},
]{hyperref}

\newcommand*{\citen}[1]{%
  \begingroup
    \romannumeral-`\x 
    \setcitestyle{numbers}%
    \cite{#1}%
  \endgroup   
}

\def\tsi{$\mathrm{(TaSe_4)_2I}$}
\begin{document}

\title{Absence of in-gap modes in charge-density-wave edge dislocations of Weyl semimetal (TaSe$_4$)$_2$I}
\author{Zengle~Huang}
\affiliation{Department of Physics $\&$ Astronomy, Rutgers
	University, Piscataway, New Jersey 08854, USA}
\author{Hemian~Yi}
\affiliation{Department of Physics, The Pennsylvania State University, University Park, Pennsylvania 16802, USA}
\author{Lujin~Min}
\affiliation{Department of Physics, The Pennsylvania State University, University Park, Pennsylvania 16802, USA}
\author{Zhiqiang~Mao}
\affiliation{Department of Physics, The Pennsylvania State University, University Park, Pennsylvania 16802, USA}
\author{Cui-Zu~Chang}
\affiliation{Department of Physics, The Pennsylvania State University, University Park, Pennsylvania 16802, USA}
\author{Weida~Wu}
\email{wdwu@physics.rutgers.edu}
\affiliation{Department of Physics $\&$ Astronomy, Rutgers	University, Piscataway, New Jersey 08854, USA}

\begin{abstract}
Axion electrodynamics in condensed matter could emerge from the formation of charge density wave~(CDW) in Weyl semimetals. (TaSe$_4$)$_2$I was proposed to be the first material platform for realizing axionic CDW that may host topological defects with 1-dimensional in-gap modes.  The real-space modulation of CDW phase and the existence of in-gap modes remain elusive. Here we present the first comprehensive scanning tunneling microscopic and spectroscopic study on the CDW on (TaSe$_4$)$_2$I (110) surface.  The tunneling spectroscopic measurements reveal a CDW gap of $\sim 200$ meV, in good agreement with prior studies,  while spectroscopy of CDW edge dislocation shows no indication of in-gap states. The bias-dependent STM images indicate that the CDW in (TaSe$_4$)$_2$I is dominated by large periodic lattice distortion instead of charge modulation,  suggesting a non-Peierls mechanism of the CDW instability.     
\end{abstract}	

\maketitle

Charge density wave~(CDW) has been a subject of long-standing research interest given its rich mechanisms~\cite{Johannes2008} and its intimate relation with novel phases of matter~\cite{Fradkin2015}. Recently, it has been shown that CDW can couple to the topological Weyl semimetals by gapping out the Weyl nodes of opposite chiralities near the Fermi energy~\cite{Wang2013,Roy2015}. The resulting insulating phase is proposed to be an axion insulator~\cite{Wang2013,Roy2015}. The sliding mode of CDW is identified as the axion and the phase of CDW is the axion angle $\theta$ in the topological term $\theta\textbf{E}\cdot\textbf{B}$. Interestingly, the  screw or edge dislocations of CDW modulations are one-dimensional (1D) states called axion strings that carry in-gap dissipationless chiral or helical modes, depending on whether the time-reversal symmetry is broken~\cite{Wieder2020, yu2020}. Recently, several theoretical studies suggest that (TaSe$_4$)$_2$I could be the first  material to host a Weyl semimetal-CDW phase transition~\cite{Shi2021, Li2021}. 

\begin{figure}[htpb]
	\includegraphics[width=\columnwidth]{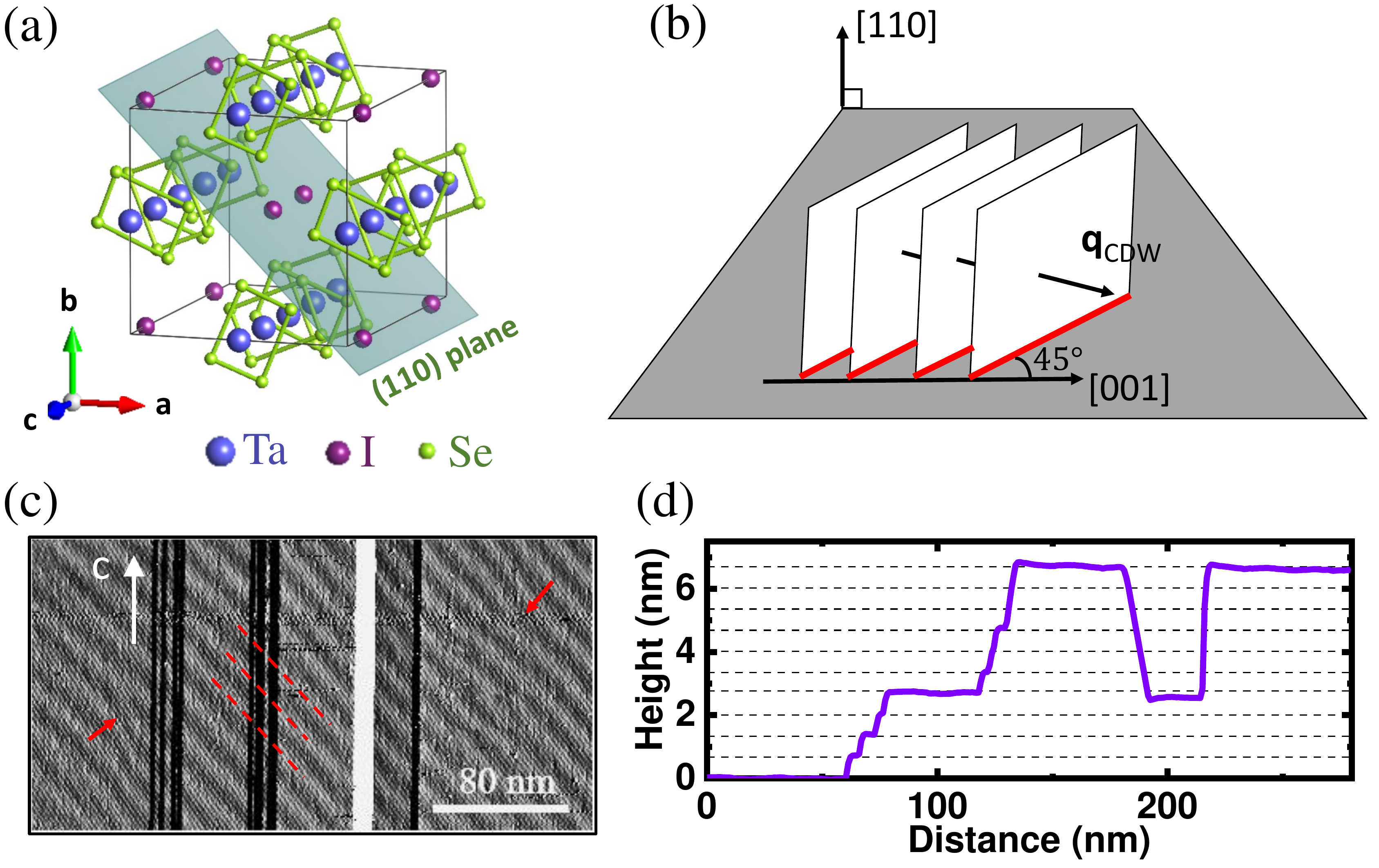}
	\caption{(Color online) (a) The conventional unit cell of (TaSe$_4$)$_2$I with the natural cleavage (110) plane in dark green. (b) A schematic of CDW orientation relative to the (110) plane and the [001] direction, for $\textbf{q}_{CDW}$ = ($+\eta$, $-\eta$, $\pm\delta$) or ($-\eta$, $+\eta$, $\pm\delta$). The white planes are the CDW wave fronts. The red lines are the intersection between the CDW wave fronts and the (110) plane. The CDW wave fronts are perpendicular to the (110) plane and the intersections are  45$^{\circ}$ to the [001] direction, corresponding to the CDW modulation observed in (c). (c) A large-scale STM topographic (derivative) image at 150~K showing multiple steps and the CDW modulation ($V$ = 1~V, $I$ = 30~pA). The step edges are along the $c$-axis because of the strong covalent bonds along the TaSe$_4$ chains. The red dashed lines are the guide for eye that track the CDW modulation across multiple steps. The red arrows highlight two CDW edge dislocations. (d) The line profile of the STM image in (c). The separation between the neighboring dash lines is the interplane distance along [110] direction. 
		\label{Fig1}     }
\end{figure}

(TaSe$_4$)$_2$I is one of the archetypal quasi-one-dimensional~(quasi-1D) CDW compounds where the emergence of spontaneous charge modulations was attributed to nesting of Fermi surfaces, the Peierls mechanism~\cite{Gruner2018}. (TaSe$_4$)$_2$I undergoes a semiconductor-insulator transition  at $T_c\approx$260~K, below which it enters an incommensurate CDW phase with modulation wave vector $\textbf{q}_{CDW}$=($\pm\eta$, $\pm\eta$, $\pm\delta$) characterized by x-ray and neutron diffraction~\cite{Fujishita1984,Lee1985,Lorenzo1998,Shi2021}. Transport measurements reveal a thermal activation gap of about 260~meV~\cite{Gressier1984,Forrr1987,Wang1983}. As shown in Fig.~\ref{Fig1}a, (TaSe$_4$)$_2$I has a body-center tetragonal unit cell which is composed of TaSe$_4$ chains  separated by iodine ions. In each TaSe$_4$ chain, Ta atoms and Se$_4$ rectangles alternate with strong covalent bonds between them, resulting in large band width along the chain direction. The Se$_4$ rectangles rotate by 45$^\circ$ relative to its neighbor along the chain and result in a chiral structure without inversion symmetry. The absence of inversion symmetry hints the possibility of hosting interesting topological electronic properties~\cite{Chang2018}. The renewed interests in (TaSe$_4$)$_2$I are motivated by the recent proposal of axionic CDW that could host topological band structure in the correlated phases~\cite{Wang2013,Roy2015}. The recent observation of positive magnetoconductance in the CDW sliding mode  provides compelling evidence of the chiral anomaly signature of axionic insulator phase in (TaSe$_4$)$_2$I~\cite{Gooth2019}.

Prior studies suggest \tsi\  is a zero-gap semiconductor above the CDW transition~\cite{Tournier-Colletta2013}. The recent first principle calculations confirms the high temperature phase of \tsi~is indeed a Weyl semimetal, one of the necessary conditions for the axionic CDW phase~\cite{Shi2021, Li2021}. The bulk band structure was confirmed by recent angle-resolved photoemission spectroscopy~(ARPES) measurements~\cite{Yi2021}. The observation that Dirac nodes along the high symmetry line is 0.4~eV below the Fermi energy~\cite{Yi2021} agrees with previous studies~\cite{Tournier-Colletta2013}. However, significant thermal broadening at room temperature prevents high resolution spectroscopy measurements of the Fermi arc surface states, the smoking-gun evidence of Weyl semimetals~\cite{Xu2015,Lv2015,Lv2015a,Inoue2016,Kim2021}.
The existence of axion strings, \textit{i.e.}, the 1D topological states at the core of topological defects of the CDW phase would be a direct evidence of the axion CDW phase~\cite{Wang2013,Roy2015}. It is challenging to carry out high resolution local spectroscopy measurements of the CDW phase because of the insulating nature of the CDW phase of \tsi\ at low temperature~\cite{Gressier1984,Forrr1987,Wang1983}. There is a lack of real-space characterization such as scanning tunneling microscopy~(STM) imaging of the CDW modulations in \tsi\ despite that the compound had been studied for decades~\cite{Fujishita1984,Lee1985,Lorenzo1998,Tournier-Colletta2013,Shi2021}. Recently, we have demonstrated the first STM imaging of the CDW phase of \tsi~\cite{Yi2021}, which opens up the possibility of tunneling spectroscopy studies of topological defects of the CDW phase.

In this work, we investigated the CDW modulation and its edge dislocations in (TaSe$_4$)$_2$I using STM and scanning tunneling spectroscopy~(STS). We observed the single-$q$ CDW modulation which agrees with the prior diffraction measurements. Our STM results reveal that the cleaved $(110)$ surface is the iodine plane with only half of the iodine ions  left on the surface to maintain charge neutrality. Tunneling spectroscopic measurements find no signature of in-gap states in the core of CDW edge dislocations within our resolution, indicating the absence of in-gap topological 1D modes.  Bias-independent CDW modulation amplitude implies that large lattice distortion dominates over a wide energy range, suggesting a strong-coupling CDW mechanism.  

(TaSe$_4$)$_2$I single crystals were grown by the chemical transport method described in Ref.~\citen{Fujishita1985}. The STM/STS results were obtained in an Omicron LT-STM with base pressure 1$\times$10$^{-11}$~mbar. Electrochemically etched tungsten tips were characterized on a clean Cu~(111) surface before STM experiments. (TaSe$_4$)$_2$I single crystals were cleaved \textit{in-situ} at room temperature and immediately inserted into the cold STM head. Scanning tunneling spectroscopy measurements were performed with standard lock-in technique with modulation frequency 899~Hz and 10~mV (root mean square).

Fig.~\ref{Fig1}(c) shows a topographic image of a freshly cleaved (110) surface of (TaSe$_4$)$_2$I, with several step edges which are parallel to the chain direction~($c$-axis). As shown in Fig.~\ref{Fig1}(d), the step heights are integer multiples of $a/\sqrt{2}$ ($\approx$0.67~nm), the interplane distance along the [110] direction. This confirms that the cleaved surface is the (110) surface~\cite{Tournier-Colletta2013}. Single-$q$ CDW modulation can be observed on the entire scanned area. The observed CDW modulation are the intersections between the three-dimensional CDW wave fronts and the (110) surface. Here the modulation wave fronts form a (45$\pm$1)$^{\circ}$ angle with the step edges, suggesting \textbf{q}$_{CDW}$=($+\eta$,$-\eta$,$\pm\delta$) or ($-\eta$, $+\eta$, $\pm\delta$). For these wave vectors, the CDW wave fronts are perpendicular to the (110) surface and their intersections form a 45$^{\circ}$ angle to the $c$-axis, as illustrated in Fig.~\ref{Fig1}(b). As shown in Fig.~\ref{Fig1}(c), there is no visible phase shift between the CDW modulations on different terraces, further confirming the CDW wave fronts are indeed perpendicular to the (110) surface. The persistence of CDW modulation across step edges also confirm that the STM observation reflects the bulk CDW modulations. For the other CDW wave vectors \textbf{q}$_{CDW}$=($+\eta$, $+\eta$, $\pm\delta$) or ($-\eta$,$-\eta$,$\pm\delta$) whose first two components have the same sign, the CDW wave fronts form a 45$^{\circ}$ angle to the (110) surface and the intersections are perpendicular to the $c$-axis. We did not observe this type of CDW domain over multiple locations on several samples with different cleaving conditions, which indicates surface prefers in-plane CDW wave vectors.  The CDW wavelength at 150~K is $\sim17\pm$1~nm, in good agreement with  recent x-ray diffraction results where $\eta$=0.027 and $\delta$=0.049~\cite{Shi2021}. However, the observed CDW wavelength is significantly longer than that in previous reports ($\sim10.6$~nm)~\cite{Fujishita1984,Lee1985,Lorenzo1998}, which might originate from different synthesis conditions.   
 
\begin{figure}[htpb]
	\includegraphics[width=\columnwidth]{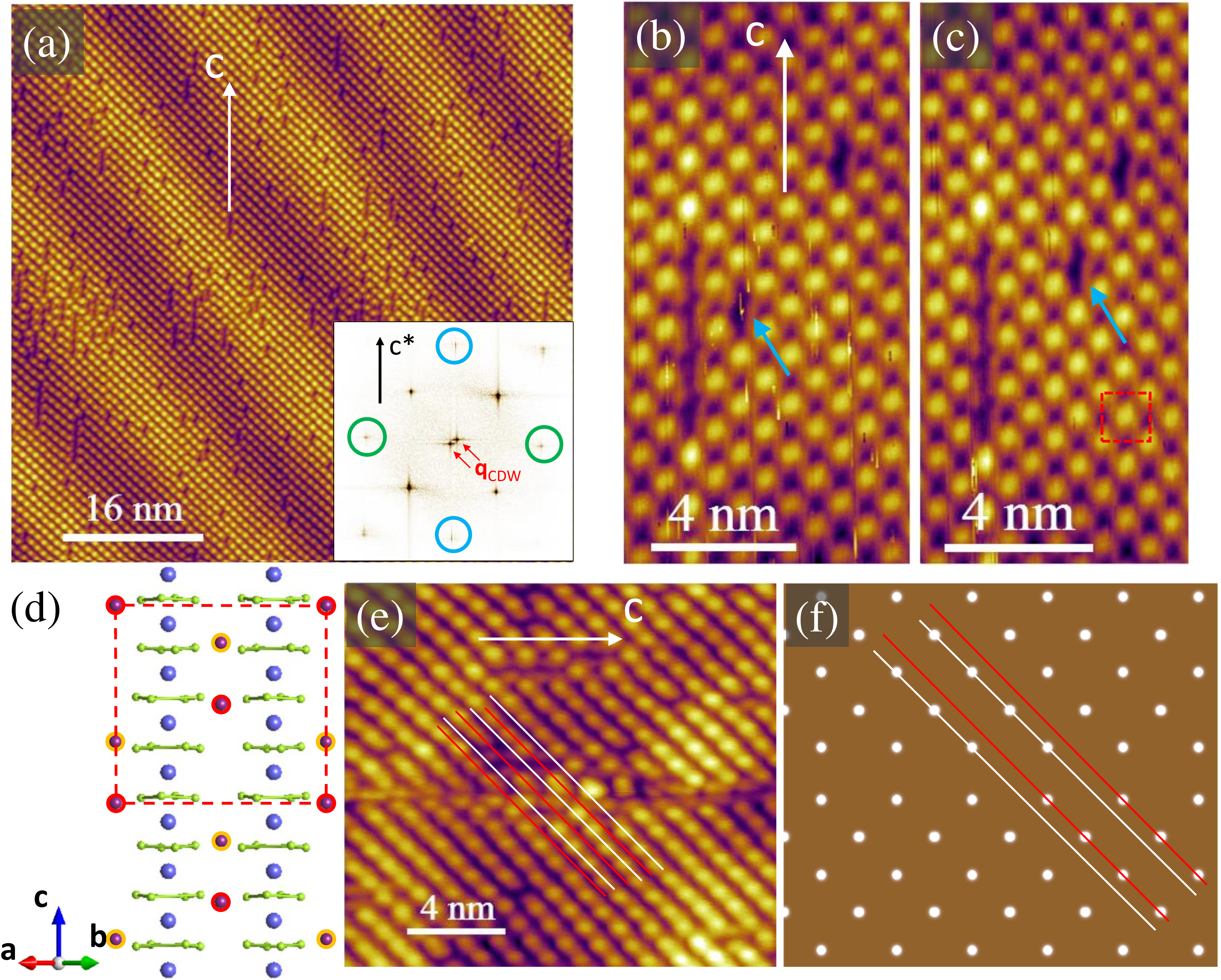}
	\caption{(Color online) (a) An atomic-resolved topographic image at 150~K showing vacancy defects on the surface ($V$ = $-$1~V, $I$ = 30~pA). The inset is the Fourier transform of the main image. The blue and green circles respectively highlight the lattice Bragg peaks used to calculate the lattice constants along and perpendicular to the c-axis. (b) and (c) are two topographic images taken consecutively ($V$ = $-$1 V, $I$ = 20~pA). The vacancy defect highlighted by the blue arrow moves during scanning. The dashed red square marks a unit cell. (d) Crystal structure of (TaSe$_4$)$_2$I projected on the (110) plane. Iodine ions with red and orange outlines are respectively on two different atomic sites. The dashed red square marks a unit cell corresponding to the one in (c). (e) A topographic image showing a crystalline antiphase boundary. The white and red lines track the lattice on both sides of the domain wall. (f) A simulated lattice of iodine ions with a domain wall separating two domains where iodine ions occupy different sites.
		\label{Fig2} 	}
\end{figure}

To identify the nature of cleaved surfaces, we obtained the atomic-resolved topographic image of cleaved (110) surface shown in Fig.~\ref{Fig2}(a). We found larger atomic corrugation at negative bias, indicating dominant occupied states on the cleaved surface. The inset of Fig.~\ref{Fig2}(a) is the Fourier transform map showing sharp CDW and lattice Bragg peaks. The CDW Bragg peaks are closed to the origin as indicated by the red arrows. The lattice Bragg peaks highlighted by the circles infer that the lattice constants are $c\approx12.67$~\AA\, and 13.98~\AA\ perpendicular to the $c$-axis, in good agreement with the reported values~\cite{Wang1983, Tournier-Colletta2013}. A considerable amount of vacancy defects can be seen on the surface. In Fig.~\ref{Fig2}(b) and \ref{Fig2}(c), two topographic images taken consecutively, the vacancy highlighted by a blue arrow moves across two atomic sites. Such defect migration at low temperature is more likely to occur on the iodine surface, where the atomic distance is larger and the ionic bonds are weaker than the covalent bonds in the $\mathrm{TaSe_4}$ chains. Therefore, the cleaved surface is likely the iodine surface and thus the vacancy defects are the iodine vacancies. Interestingly, some iodine vacancies tend to form vacancy chains along $c$-axis, as shown on the left of Fig.~\ref{Fig2}(b) and \ref{Fig2}(c), probably driven by anisotropic screening of the quasi-1D electronic structure. 

However, there are two different atomic sites for iodine ions within one unit cell as shown in Fig.~\ref{Fig2}(d). We only observe one iodine per surface unit cell, suggesting 50\% of iodines are missing after cleaving. Because iodine ions are negatively charged, naturally half of iodine ions would stay on one of the surfaces upon cleaving to maintain  charge neutrality. In addition, they would collectively occupy one of the two sites to minimize Coulomb interaction.  Thus, there would be two antiphase domains on cleave surfaces. Our atomic-resolved topographic images confirm that this is indeed the case. Fig.~\ref{Fig2}(e) shows an example of an antiphase boundary separating two domains across which there is a relative phase shift highlighted by the straight lines. In the simulated lattice shown in Fig.~\ref{Fig2}(f), a domain wall separates two domains with iodine ions on different atomic sites, resulting in the same phase shift. Using Fourier analysis, we confirm the numerical values of the phase shift in the STM image and the simulation are the same,  approximately 0.64$\pi$  across the antiphase domain wall  (see Fig.~S1 in Supplemental Material~\cite{SM}). Therefore, our STM results and analysis provide clear evidence that the cleave surfaces in \tsi\ are iodine plane with 50\% occupancy. The density of iodine ions on the surface determines the surface electric potential and  band bending near surface. Recent ARPES measurements found that the loss of surface iodine ions due to thermal annealing introduces a substantial surface electric field that induces a Rashba-like splitting of the Dirac bands~\cite{Yi2021}.              

\begin{figure*}
	\includegraphics[width=\textwidth]{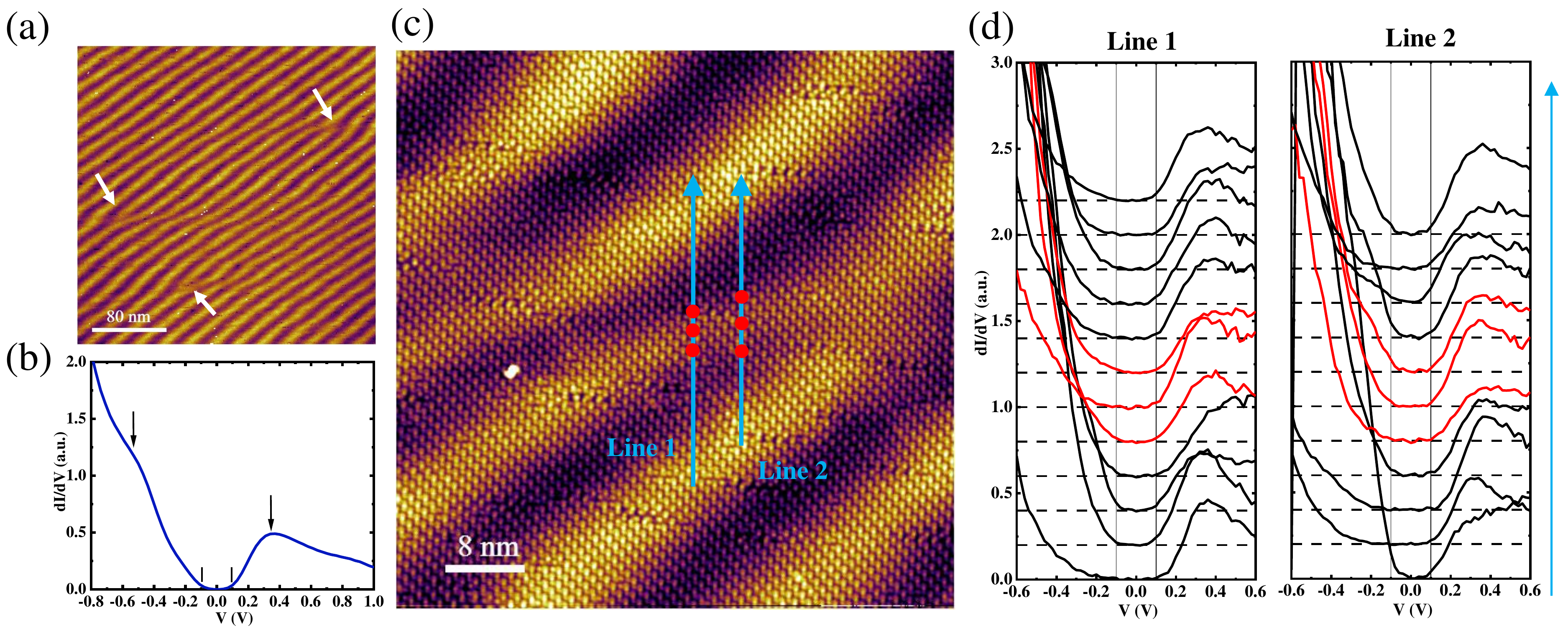}
	\caption{(Color online) (a) A topographic image at 150~K showing three CDW edge dislocations on the surface. Three white arrows highlight the CDW edge dislocations. (b) The averaged tunneling spectrum at 150~K of regions without dislocation ($V$=1~V, $I$=200~pA). Arrows highlight two reproducible features in the tunneling spectra: a shoulder at about $-$0.5~V and a broad peak at about 0.3~V. A gap of around 200~meV is indicated by two solid lines. (c) A topographic image with an edge dislocation. The tunneling spectra in (d) are taken along the two solid arrows. The red spectra in (d) are taken on the red dots. (d) Tunneling spectra taken along the two solid lines in (c). The spectra are offset for clarity. The dashed lines indicate the zero in the spectra. The red curves are the spectra taken on the edge dislocation. ($V$=0.6~V, $I$=100~pA). The solid lines indicates the approximate position of the gap.  
		\label{Fig3} 	}
\end{figure*}

The quasi-1D system (TaSe$_4$)$_2$I is proposed to host the axionic CDW, which requires the nesting between two Weyl nodes of opposite chirality near Fermi energy~\cite{Gooth2019}. It has been predicted that a signature of axionic CDW is the 1D helical or chiral modes propagating along the CDW edge or screw dislocation with Burger's vector $-\frac{2\pi \textbf{q}}{|\textbf{q}|^2}$, where $\textbf{q}$ is the CDW wave vector. The origin of the 1D modes closely resemble the edge states of a 2D quantum Hall system which cross the Fermi level inside the band gap~\cite{Wang2013,Roy2015}.The 1D modes would manifest as in-gap states in the single-particle spectra, providing smoking-gun evidence for axionic CDW phase. CDW edge dislocation is ubiquitous on (TaSe$_4$)$_2$I surface as shown in Fig.~\ref{Fig3}(a), where three CDW edge dislocations are indicated by the white arrows. Fig.~\ref{Fig3}(b) shows the averaged tunneling spectra at the areas without CDW dislocations. A shoulder-like feature appears around $-$0.5~V in the valence band and a broad peak is around 0.3~V in the conduction band. A gap of about 200~meV centers around Fermi energy, in good agreement with prior transport measurements~\cite{Gressier1984,Forrr1987,Wang1983,Gooth2019}. Theses three features are reproducible with different tunneling setpoints, tips, and  samples,  confirming that the observed energy gap is intrinsic. In Fig.~\ref{Fig3}(d), we plot the spectra taken along two lines crossing the CDW edge dislocation in Fig.~\ref{Fig3}(c). For the red spectra taken on the center of edge dislocation, local density of states~(LDOS) inside the gap remain the same as other spectra taken outside the dislocation core. There are some spatial fluctuations of the valence band DOS and the position of the broad peak in the conduction band, but they are not associated with the CDW edge dislocation and are likely to be related to the short arrays of iodine vacancies seen in topography (see Fig. S2 in Supplemental Material~\cite{SM}). Therefore, no signature of 1D modes at the CDW edge dislocation was found from the local tunneling spectra within our resolution. It is possible that the spectral weight of the 1D modes is too small to be resolved by our STS measurements.  On the other hand, the 1D modes could be absent for several reasons. While low-energy field theory links Weyl semimetal-CDW  phase to axion insulators~\cite{Wang2013,Roy2015}, an analytic and numerical study that considers bands at higher energy finds that magnetic Weyl-CDW phases are ``obstructed'' quantum anomalous Hall~(QAH) insulators~\cite{Wieder2020}. Alternatively, the CDW gap size might exceed the band inversion due to spin-orbital coupling, resulting in a trivial band structure. Last but not the least, the defect-induced carrier doping could shift the Weyl nodes away from Fermi energy~\cite{Yi2021,Tournier-Colletta2013}.  Thus, further studies are needed to clarify the topological nature of the CDW phase in \tsi. 
     
To gain more insight into the mechanism of CDW in (TaSe$_4$)$_2$I, we perform bias-dependent STM measurements as shown in Fig.~\ref{Fig4}(a-f) (see Fig.~S3 in Supplemental Material~\cite{SM} for more biases). In the purely electronic Peierls picture, CDW modulation is dominated by the charge modulation because of nesting of Fermi surface. The lattice weakly distorts in response to the charge modulation because of the elastic energy cost~\cite{Gruner2018}. The charge modulation also results in the periodic modulation of LDOS so that the densities of empty and occupied states vary with the same period but out of phase~\cite{Spera2020,Dai2014}. Therefore, in the constant-current topographic images, contrast inversion is expected between the negative~(occupied states) and the positive~(unoccupied states) biases.  In Fig.~\ref{Fig4}(a), the lower half of the topographic image was obtained at $+$1~V, while the upper half was taken at $-$1~V. Clearly there is no contrast inversion of the CDW modulation: its crest and trough remain at the same phase as illustrated in Fig.~\ref{Fig4}(g). The absence of contrast inversion in \tsi\ is qualitatively different from the systems with Peierls-instability-driven CDW, such as NbSe$_3$~\cite{Brun2009}, purple bronze~\cite{Mallet1999}, and Pb/Ge(111)-$\alpha$~\cite{Carpinelli1996} where charge modulation dominates, suggesting a non-Periels CDW mechanism in \tsi.

Furthermore,  the electronic modulation due to Fermi surface nesting would be mostly pronounced around the CDW gap energy in the Peierls picture as sketched in Fig.~\ref{Fig4}(h)~\cite{Spera2020,Dai2014}. This was demonstrated in the CDW phase of chromium~\cite{Braun2000}. We extract in Fig.~\ref{Fig4}(h) the bias-dependent amplitudes at 150~K using the sinusoidal fits of line profiles shown in Fig.~\ref{Fig4}(g). The modulation amplitude ($\sim$20 to 25~pm) is almost independent of tunneling bias at 150~K at various biases within $\pm$1~V. We cannot obtain reliable topography data for low biases $|V|<$0.6~V with practical tunneling junction resistance, likely due to poor bulk conduction of the sample. Nevertheless, the tunneling spectroscopic maps in this range do not reveal any LDOS modulation associated to the CDW (see Fig.~S2 in Supplemental Material~\cite{SM}). The bias-independence of modulation amplitude suggests that the CDW in \tsi\ is dominated by the periodic lattice distortion instead of the charge modulation. 

In addition, the observed lattice distortion ($>20$~pm) is $\sim2\%$ of lattice constant, and is perpendicular to the CDW wave vector~\cite{Lee1985}. The large transverse lattice distortion is inconsistent with simple Periels picture which favors weak longitudinal distortion. On the other hand, the significant lattice distortion in \tsi\ is comparable with two-dimensional transition metal dichalcogenides such as 1$T$-TaS$_2$, 2$H$-TaSe$_2$ and 1$T$-TiSe$_2$, where CDW are likely driven by the strong-coupling mechanism though the issue is not completely settled~\cite{Rossnagel2011, Spera2020}. Therefore, although (TaSe$_4$)$_2$I is a quasi-1D compound, our results suggest that its CDW phase may be driven by a non-Peierls mechanism.

We also performed STM measurements at 110, 260 and 300~K, as shown in Fig.~\ref{Fig5}. The line profiles in Fig.~\ref{Fig5}(e) indicate that the CDW modulation becomes weaker as the temperature approaches $T_\mathrm{C}=263$~K and disappears at 300~K. The temperature dependence follows the mean-field behavior as shown in Fig.~\ref{Fig5}(f), in good agreement with prior studies~\cite{Wang1983, Fujishita1984,Gressier1984,Forrr1987, Lee1985,Lorenzo1998}. This further confirms our STM observation reflects the intrinsic properties of the bulk CDW phase in \tsi.

In summary, the CDW modulation and edge dislocations in \tsi\ are characterized STM/STS.  
We cannot resolve any in-gap states at the core of edge dislocations, and thus cannot confirm the axionic nature of CDW phase in Weyl semimetal (TaSe$_4$)$_2$I. The CDW modulation in (TaSe$_4$)$_2$I is dominated by large periodic lattice distortion with negligible electronic contribution, suggesting that the CDW phase is primarily driven by lattice-related mechanism instead of the conventional Peierls one. Our results  not only  will inspire further theoretical investigations and scanning probe experiments to clarify the mechanism and topological nature of the CDW phase in (TaSe$_4$)$_2$I, but also will stimulate further exploration of the 1D in-gap modes in Weyl-CDW systems.   	      

We thank Benjamin Wieder and David Vanderbilt for fruitful discussions. The STM at Rutgers was supported by the ARO Award (Grant No. W911NF-20-1-0108). The crystal growth (L.J.M. and Z.Q.M.) was supported by a Penn State NSF-MRSEC Grant (Grant No. DMR2011839). C.Z.C acknowledges the support from the NSF-CAREER Award (Grant No. DMR-1847811), the Gordon and Betty Moore Foundation’s EPiQS Initiative (Grant No. GBMF9063 to C.-Z.C.) and a Penn State NSF-MRSEC Grant (Grant No. DMR-2011839).

\bibliography{Ta2Se8I}

\begin{figure*}[htpb]
	\includegraphics[width=\columnwidth]{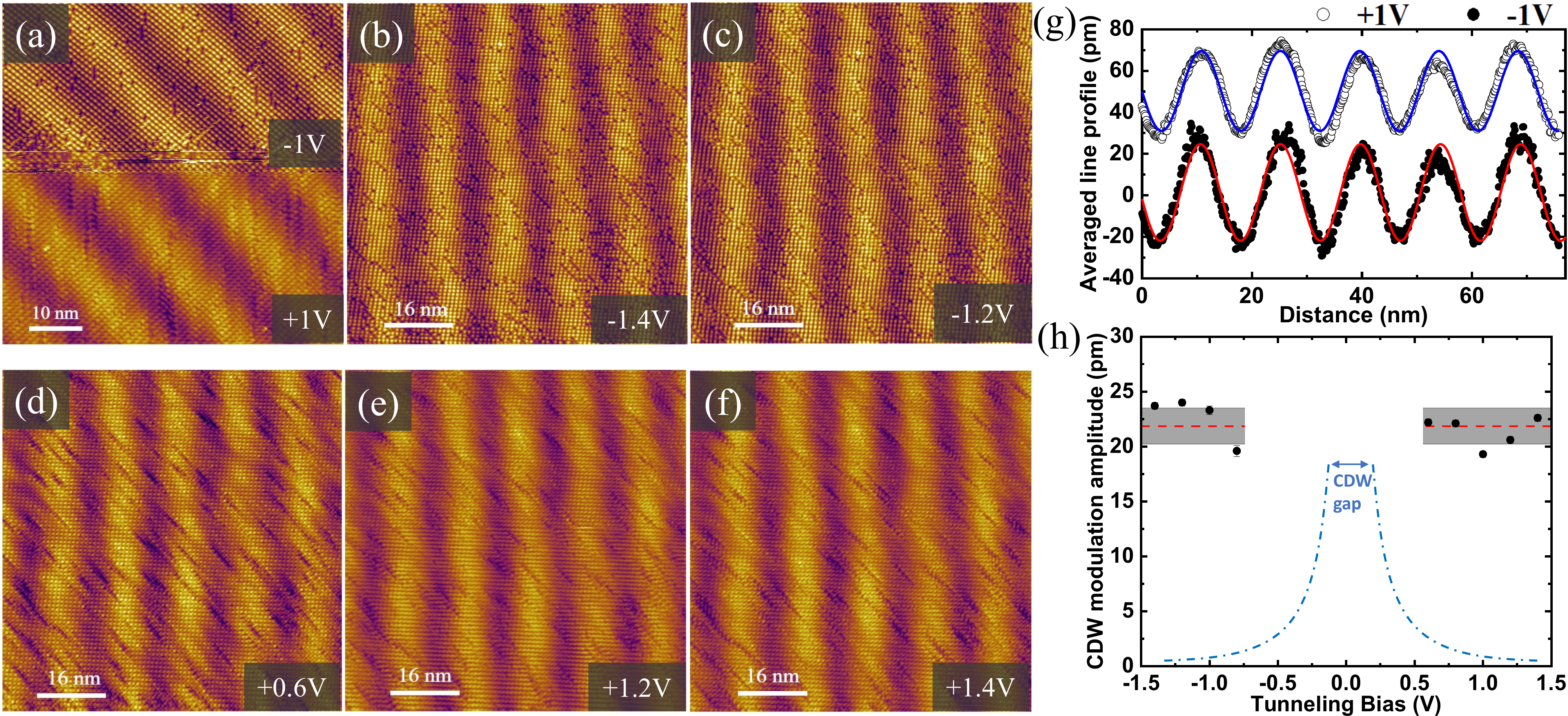}
	\caption{(Color online) (a) A topographic image at 150~K with the lower (upper) half taken at +1~V ($-1$~V) bias. The current setpoint is 30~pA. (b-f) Bias-dependent topographic images of CDW modulation. The current setpoint is 20~pA for all images. The color scales are all $-$60 to 60~pm. (g) The averaged line profile of CDW modulation at $-1$~V (black dots) and $+1$~V (white dots). The blue and red curves are the sinusoidal fits of the line profiles. The $+$1~V data points are shifted upwards by 50~pm for clarity. (h) CDW modulation amplitude versus tunneling bias. All topographic images where the amplitudes are obtained are shown in Fig.~S3 in the Supplemental Material~\cite{SM}. The amplitudes are obtained from the sinusoidal fits. The red dashed line and the gray band are the mean and standard deviation of the amplitudes. The blue dash-dotted line is a sketch of the expected bias dependence of the modulation amplitude for a Peierls-driven CDW.
		\label{Fig4} 	}
\end{figure*}

\begin{figure}[htpb]
	\includegraphics[width=\columnwidth]{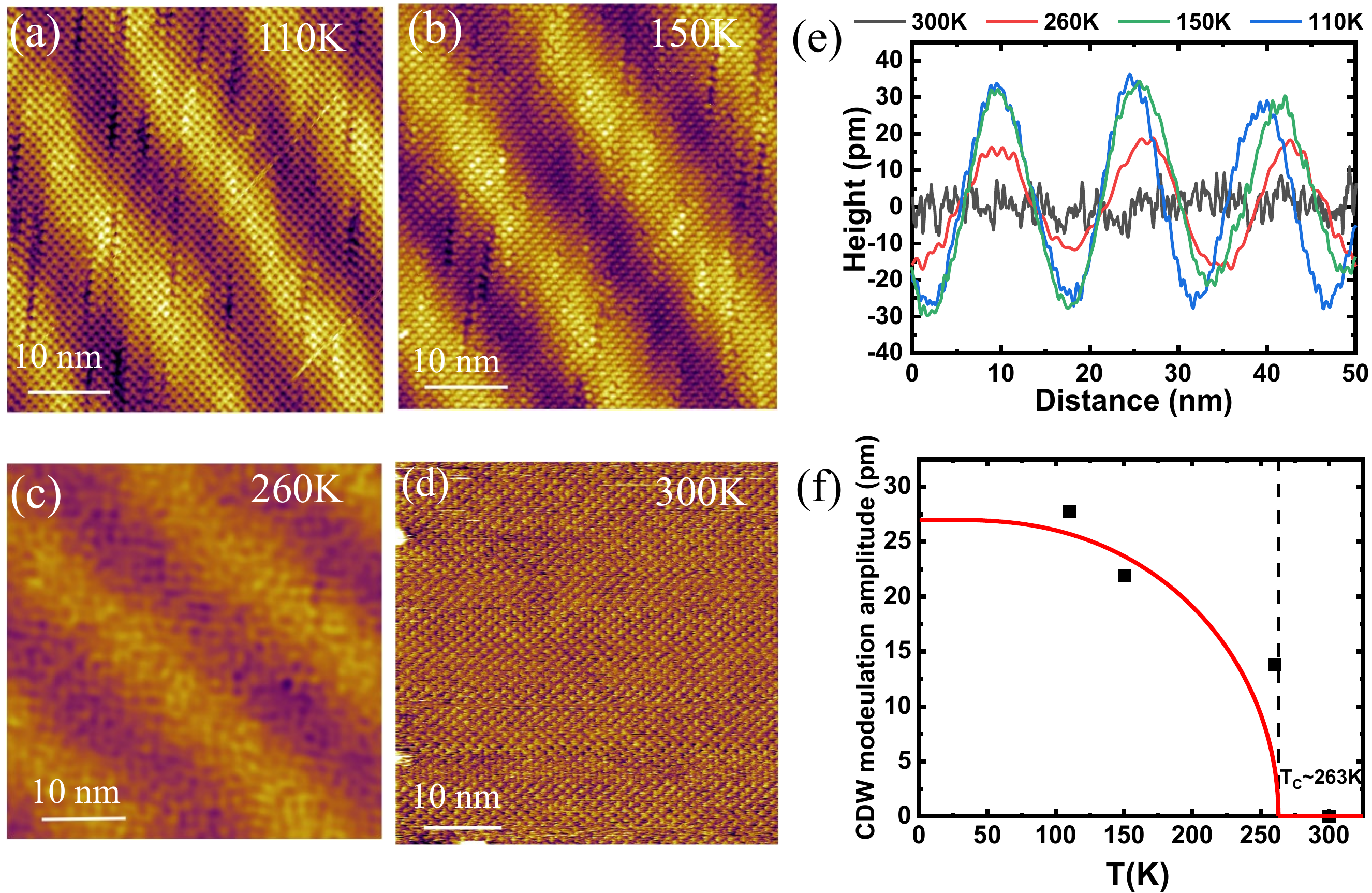}
	\caption{(Color online) (a-d) Topographic images of CDW modulation at 110~K, 150~K, 260~K and 300~K. ($V$=$+$1~V, $I$=30~pA for (a-c); $V$= $-$1~V, $I$=30~pA for (d)). The color scales are all $-$50 to 50~pm. (e) Temperature-dependent CDW line profiles obtained from (a-d). (f) Temperature-dependent CDW modulation amplitude. The error bar is smaller than the size of the data points. The red line is the guide for eye.
		\label{Fig5} 	}
\end{figure}

\end{document}